\documentclass[12pt,psfig]{article}
\usepackage{amssymb}
\usepackage{amsmath,amsthm}
\usepackage{amsfonts}
\usepackage{graphicx}
\usepackage{graphics}
\numberwithin{equation}{section}

\usepackage{hyperref}

\begin{document}
\begin{center}\Large\textbf{Interaction of the 
branes in the Presence of the Background Fields: the
Dynamical Non-intersecting Perpendicular Wrapped-Fractional
Configuration}
\end{center}
\vspace{0.75cm}
\begin{center}{\large Elham Maghsoodi and \large Davoud
Kamani}\end{center}
\begin{center}
\textsl{\small{Physics Department, Amirkabir University of
Technology (Tehran Polytechnic)\\
P.O.Box: 15875-4413, Tehran, Iran\\
e-mails: kamani@aut.ac.ir , el.maghsoodi@aut.ac.ir\\
}}
\end{center}
\vspace{0.5cm}

\begin{abstract}

We shall obtain the interaction 
of the D$p_1$- and D$p_2$-branes 
in the toroidal-orbifold spacetime 
$\mathit{T}^{n}\times\mathbb{R}^{1, d-n-5} 
\times\mathbb{C}^{2}/\mathbb{Z}_{2}$.
The configuration of the branes is: non-intersecting, perpendicular, 
moving-rotating, wrapped-fractional with background fields. 
For this, we calculate the bosonic boundary state corresponding to a
dynamical fractional-wrapped D$p$-brane 
in presence of the Kalb-Ramond
field, a $U(1)$ gauge potential and an open string tachyon field.
The long-range behavior of the interaction amplitude 
will be extracted.

\end{abstract}


\newpage
\section{Introduction}

In development of string theory, a crucial 
role is played by D-branes \cite{1}. 
An adequate tool for describing D-branes and
their interactions is the boundary state formalism 
\cite{2}-\cite{20}. 
In this method a D-brane appears as a source (sink) for 
emitting (absorbing) all closed string states. 
All properties of a D-brane are prominently encoded on a 
boundary state, hence, the interaction 
between two D-branes is acquired  
by the overlap of their corresponding boundary states via
the closed string propagator.
Thus, the boundary state approach for various setups
of the stationary and dynamical  
D-branes with internal background fields in
the compact and non-compact spacetimes has been applied
\cite{9}-\cite{15}.
In fact, among these configurations the systems with 
fractional D-branes exhibit some appealing and marvelous 
behaviors \cite{16}-\cite{23}. 

Therefore, in one hand we have the fractional branes which appear 
in the various parts of the M- and string theories. 
For example, propagating fractional branes provide a 
simple and explicit starting point for the definition 
of Matrix theory \cite{24,25}.
Besides, they have been used for demonstrating the 
gauge/gravity correspondence \cite{21}. 
On the other hand, we have the wrapped branes which 
have a widespread application in string theory. 
The fractional and wrapped branes motivated us to 
study a special setup of the fractional-wrapped branes. 
In our definition 
these are the wrapped branes which live in the fixed 
points of the non-compact orbifold $\mathbb{C}^{2}/\mathbb{Z}_{2}$. 

In this paper we apply the boundary state formalism to 
calculate the interaction amplitude between two 
non-intersecting perpendicular  
fractional-wrapped D-branes in the framework 
of the bosonic string theory. As a result, 
we shall demonstrate that intersections of the worldsheet 
of the exchanged closed string 
with the worldvolumes of the branes create the interaction.  
However, the background spacetime partially is compact on a torus,
and has the topological structure 
\begin{eqnarray}
T^n \times \mathbb{R}^{1, d-n-5} 
\times\mathbb{C}^{2}/\mathbb{Z}_{2}\;\;,\;\;
n \in \{0,1,\ldots,d-5\}.
\nonumber
\end{eqnarray}
We shall consider an arbitrary torus of the set
$\{T^n|n =0,1,\ldots,d-5\}$. 
In addition, we introduce the antisymmetric tensor $B_{\mu\nu}$, 
the $U(1)$ gauge potentials   
and open string tachyon fields on the worldvolumes of the branes.
The branes are dynamical, i.e. they are rotating and 
are moving within their volumes. In this setup 
there are various parameters 
which the strength of the interaction can be adjusted by them.  
We shall show that the long-range force, 
extracted from the interaction amplitude, 
has a damping nature. 

Note that the complete form of the theory includes 
both the twisted and untwisted sectors under the 
$\mathbb{Z}_2$-group. Thus, the total interaction 
amplitude is sum of the amplitudes of the twisted and 
untwisted sectors. However, since for the various 
setups the untwisted sector has been extremely studied 
we shall only concentrate on the twisted sector of 
our setup. 

This paper is organized as follows. In Sec. 2, we obtain 
the boundary state associated with a 
dynamical fractional-wrapped D$p$-brane 
with the above-mentioned internal background fields. 
In Sec. 3, the interaction amplitude of two 
non-intersecting perpendicular D$p_1$- and D$p_2$-branes
will be calculated. In Sec. 3.1, behavior of the amplitude 
concerning the large distances of the branes will be investigated. 
Section 4 is devoted to the conclusions.

\section{The D$p$-brane boundary state}
\hspace{0.5cm}
Our starting point is to consider a fractional 
D$p$-brane which lives in the $d$-dimensional
spacetime, including a toroidal portion 
$T^n$, and an orbifold part
$\mathbb{C}^{2}/\mathbb{Z}_{2}$
where the $\mathbb{Z}_2$ group
acts on the coordinates $\{x^a|a= d-4, d-3, d-2, d-1\}$. The
orbifold is noncompact, so its fixed points are located at 
the hyperplane $x^a=0$. The D$p$-brane is 
stuck at these fixed points. 
In the $d$-dimensional orbifoldized spacetime
the brane can possesses the dimensions
$p \leq d-5$. 

We begin with the following sigma-model action for closed string
\begin{eqnarray}
S=&-&\frac{1}{4\pi \alpha'}\int_{\Sigma}d^2\sigma
\left(\sqrt{-g}g^{ab}G_{\mu\nu}\partial_{a}X^{\mu}
\partial_b X^{\nu}+\epsilon^{ab}B_{\mu\nu}\partial_a X^{\mu}
\partial_b X^{\nu}\right)
\nonumber\\
&+&\frac{1}{2\pi \alpha'}\int_{\partial\Sigma}d\sigma\left(A_{\alpha}
\partial_{\sigma}X^{\alpha}+\omega_{\alpha\beta}J^{\alpha\beta}_{\tau}+
T^2 (X)\right).
\end{eqnarray}
This action contains the antisymmetric field $B_{\mu\nu}$, 
the tachyon field $T^2(X)$ and the $U(1)$ 
gauge potential $A_\alpha(X)$. Since the states of the 
tachyon and gauge fields belong to the open 
string spectrum they intrinsically adhere to the brane,
and hence their corresponding fields 
appear in the boundary action.  
$g^{ab}$ and $G_{\mu\nu}$ 
are the metrics of the worldsheet and the $d$-dimensional 
spacetime, respectively. $\Sigma$ is the worldsheet of
the emitted closed string, and $\partial\Sigma$ is its boundary. 
The set $\{x^{\alpha}\}$ specifies the directions 
along the D$p$-brane worldvolume.

Here we assume the background fields $G_{\mu\nu}$ 
and $B_{\mu\nu}$ to be constant, and we apply 
$T^{2}= \frac{1}{2}U_{\alpha\beta}X^{\alpha}X^{\beta}$
as the profile of the tachyon field
with the constant symmetric matrix $U$. For the $U(1)$ 
gauge potential we use the trusty gauge 
$A_{\alpha}=-\frac{1}{2}F_{\alpha \beta }X^{\beta}$ where
the field strength is constant.
The constant antisymmetric angular velocity 
$\omega_{\alpha\beta}$ represents the linear motion
and rotation of the brane, and 
$J^{\alpha \beta}_\tau=X^\alpha \partial_\tau X^\beta
-X^\beta \partial_\tau X^\alpha$
specifies the angular momentum density. Therefore, the 
dynamics of the brane is inside its volume.
In fact, the internal fields draw some specific alignments in 
the brane, and hence in the brane worldvolume
there is a broken Lorentz symmetry. 
This implies that this dynamics is imaginable.

Vanishing of the variation of this action with respect to 
${X^{\mu}(\sigma,\tau)}$ gives
the equation of motion and the following 
boundary state equations
\begin{eqnarray}
&~&\left(\mathcal{Q}_{\alpha\beta}\partial_{\tau}X^{\beta}
+\mathcal{F}_{\alpha\beta}\partial_\sigma
X^{\beta}+U_{\alpha\beta}
X^\beta+B_{\alpha i}\partial_\sigma
X^i + B_{\alpha a}\partial_\sigma
X^a\right)_{\tau=0}|B_x\rangle=0~,
\nonumber\\
&~&\left(X^i-y^i\right)_{\tau=0}|B_x\rangle=0~,
\nonumber\\
&~&\left(X^a-y^a\right)_{\tau=0}|B_x\rangle=0~,
\end{eqnarray}
where the total field strength possesses the definition
$\mathcal{F}_{\alpha\beta}= B_{\alpha\beta}-F_{\alpha\beta}$, and
we defined $\mathcal{Q}_{\alpha\beta}=
\eta_{\alpha\beta}+4\omega_{\alpha\beta}$.
The coordinates $\{x^i\}$ refer
to the non-orbifoldy directions,
perpendicular to the brane worldvolume. The
parameters $\{y^i\}$ 
specify the location of the brane. 
Due to the location of the brane at the 
fixed points of the orbifold we have 
$\{y^a =0 |a = d-4, \cdot\cdot\cdot,d-1\}$.

The mode expansion of the closed string coordinates along the
non-orbifoldy directions $x^\alpha$ and $x^i$ is
\begin{equation}
X^\lambda(\sigma,\tau)=x^\lambda+2\alpha'p^\lambda
\tau+2L^{\lambda}\sigma+\frac{i}{2}\sqrt{2\alpha'}\sum_{m\neq0}\frac{1}{m}
\left(\alpha_m^{\lambda}e^{-2im(\tau- \sigma)}+\tilde{\alpha}_m^\lambda
e^{-2im(\tau+\sigma)}\right)~,\;\lambda \in \{\alpha , i\},
\end{equation}
where $L^{\lambda}$ is zero for the non-compact directions and 
$L^{\lambda}=N^{\lambda}R^{\lambda}$ for the compact directions.
$R^{\lambda}$ is the radius of compactification and 
$N^{\lambda}$ is the winding number
of the closed string around the compact direction $x^{\lambda}$.
Note that the extra dimensions, which are required by 
string theory, are compact so small that they cannot be observed.
Hence, we introduced the toroidal compactification 
on some spatial directions.  
However, the closed string coordinates along the orbifoldy directions
have the following solution   
\begin{eqnarray}
X^a(\sigma,\tau)=\frac{i}{2}\sqrt{2\alpha'}
\sum_{r\in\mathbb{Z}+1/2}
\frac{1}{r}\left(\alpha_r^{a}e^{-2ir(\tau- \sigma)}+\tilde{\alpha}_r^a
e^{-2ir(\tau+\sigma)}\right).
\end{eqnarray}

Using the above mode expansions we acquire   
\begin{eqnarray}
&~&\left(2\alpha'\mathcal{Q}_{\alpha\beta}p^{\beta}
+2\mathcal{F}_{\alpha\beta}
L^{\beta}+U_{\alpha\beta}x^{\beta}
\right)|B\rangle^{(0)}=0~,
\nonumber\\
&~& U_{\alpha\beta}L^{\beta}|B\rangle^{(0)}=0~,
\nonumber\\
&~& (x^{i}-y^{i})|B\rangle^{(0)}=0~,
\nonumber\\
&~&L^{i}|B\rangle^{(0)}=0~,
\end{eqnarray}
for the zero-mode portion of Eqs. (2.2), and 
\begin{eqnarray}
&~&\left[\left(\mathcal{Q}_{\alpha\beta}
-\mathcal{F}_{\alpha\beta}+\dfrac{i}{2m}
U_{\alpha\beta}\right)
\alpha_{m}^{\beta}+\left(\mathcal{Q}_{\alpha\beta}
+\mathcal{F}_{\alpha\beta}
-\dfrac{i}{2m}U_{\alpha\beta}\right)
\tilde{\alpha}_{-m}^{\beta}
\right]|B\rangle^{\rm (osc)}=0~,
\nonumber\\
&~&(\alpha_{m}^{i}- \tilde{\alpha}_{-m}^{i})|B\rangle^{\rm (osc)}=0~,
\nonumber\\
&~&(\alpha_{r}^{a}-\tilde{\alpha}_{-r}^{a})|B\rangle^{\rm (osc)}=0~,
\end{eqnarray}
for the oscillating part of Eqs. (2.2). 
The second equation of Eqs. (2.5) implies that 
for an invertible matrix $U_{\alpha \beta}$
we receive $\ell^{\alpha}=0$, where $\ell^{\alpha}$
is eigenvalue of $L^{\alpha}$.
Thus, the tachyon field wonderfully prevents  
the closed strings from winding around the 
compact directions of the brane.
For a non-invertible matrix $U_{\alpha\beta}$   
the closed strings are permitted to possess 
such windings. However, for the next purposes 
we assume that $U_{\alpha\beta}$ to be invertible.
The last equation of Eqs. (2.5) reveals  
that if the direction $x^i$ is compact we obtain $\ell^{i}= 0$, 
consequently, the closed strings cannot wind around 
the non-orbifoldy perpendicular directions.

The first equation of Eqs. (2.5) induces the relation
\begin{eqnarray}
p^{\alpha} =-\frac{1}{2\alpha'}
[(\mathcal{Q}^{-1}U)^{\alpha}_{\;\;{\beta}}x^{\beta}+
2(\mathcal{Q}^{-1}\mathcal{F})^{\alpha}_{\;\;{\beta}} 
\ell^{\beta}].
\end{eqnarray}
That is, the momentum of an emitted closed string,
along the brane worldvolume,
depends on its center of mass position, the rotation
and motion of the brane, the background fields and 
its winding numbers around the compact directions of the brane.
This momentum clarifies that a peculiar potential 
drastically acts on the emitted closed strings.

Using the coherent state method, the oscillating part of the
boundary state takes the form 
\begin{eqnarray}
|B\rangle^{\rm (osc)} &=&\prod_{n=1}^{\infty}[\det{M_{(n)}}]^{-1}
\exp\left[{-\sum_{m=1}^{\infty}
\left(\frac{1}{m}\alpha_{-m}^{\lambda}S_{(m)\lambda\lambda'}
\tilde{\alpha}_{-m}^{\lambda'}\right)}\right]\nonumber\\
\nonumber\\
&\times&\exp\left[-\sum_{r=1/2}^{\infty}
\left(\frac{1}{r}\alpha_{-r}^{a}\tilde{\alpha}_{-r}^{a}\right)\right]
|0\rangle_\alpha \otimes|0\rangle_{\tilde{\alpha}}~,\label{aos}
\end{eqnarray}
where $\lambda,\lambda'\in\{\alpha,i\}$.
The matrix $S_{(m)}$ is given by 
\begin{eqnarray}
S_{(m)\lambda\lambda'}&=&\left((M_{(m)}^{-1}
N_{(m)})_{\alpha\beta},-\delta_{ij}\right)~,
\nonumber\\
M_{(m)\alpha\beta}&=&\mathcal{Q}
_{\alpha\beta}-\mathcal{F}_{\alpha\beta}
+\dfrac{i}{2m}U_{\alpha\beta}~,
\nonumber\\
N_{(m)\alpha\beta}&=&\mathcal{Q}_{\alpha\beta}
+\mathcal{F}_{\alpha\beta}-\dfrac{i}{2m}U_{\alpha\beta}~.
\end{eqnarray}
Advent of the normalization factor is 
anticipated by the disk partition function \cite{26}-\cite{28}.
In fact, the coherent state method gives the boundary state (2.8)
under the constraint $S_{(m)}S_{(-m)}^{T}=\mathbf{1}$.
This equation introduces some relations among the parameters
$\{\omega_{\alpha\beta},U_{\alpha\beta},\mathcal{F}_{\alpha\beta}\}$,
and hence reduces the number of independent parameters.

The boundary state associated with the zero modes 
possesses the following solution
\begin{eqnarray}
|B\rangle^{(0)}&=&\frac{T_p}{2\sqrt{\det(U/2)}}
\int_{-\infty}^{\infty}
\exp\left[\frac{i}{2}\alpha' \left(U^{-1}\mathcal{Q}
+\mathcal{Q}^{T}U^{-1}\right)_{\alpha\beta}\;
p^{\alpha}p^{\beta}+i(U^{-1}\mathcal{F})_{\alpha\beta}
\ell^{\alpha}p^{\beta}
\right]
\nonumber\\
&\times & \prod_{\alpha=0}^{p}\left[|p^{\alpha}\rangle 
dp^{\alpha}\right]\otimes \prod_{i=p+1}^{d-5}
\left[ \delta \left({x}^{i}-y^{i}\right)|p^{i}_{L}=p^{i}_{R}=0\rangle
\right]~
\label{zer}.
\end{eqnarray}
The total boundary state is
$$|B\rangle=|B\rangle^{\rm (osc)} \otimes|B\rangle^{(0)}
\otimes|B_{\rm gh}\rangle~,$$
where $|B_{\rm gh}\rangle$ is the known boundary 
state corresponding to the conformal ghost fields
\begin{equation}
|B_{\rm gh}\rangle=\exp{\left[\sum_{m=1}^{\infty}(c_{-m}\tilde{b}_{-m}
-b_{-m} \tilde{c}_{-m})\right]}\frac{c_0+\tilde{c}_0}{2}
|q=1\rangle|\tilde{q}=1\rangle~.
\end{equation}
This boundary state is independent of 
the orbifold projection, brane dynamics, compactification and the
background fields.
\section{The D-branes interaction}

For showing the importance of 
the D-branes interactions and connection of 
this subject with the main problems of physics, 
for example, one can say that
such interactions in the brane-world have
been proposed as origin of the  
inflation \cite{29, 30}. In addition,
Big-Bang has been actually created by 
the collision of two D-branes \cite{31}.
These interactions also have been 
considered in the early universe
for describing the radiation-dominated era.
Besides, since the gravity is a force 
that can penetrate and interact across the branes, 
interaction of these hypersurfaces 
produce the added gravity inside our D3-brane,
i.e. our world \cite{32, 33}. Furthermore, the dark matter 
somehow can be a result of the gravitational interaction 
between the branes \cite{34}. 
Finally, the branes interactions also shed light on the 
gauge/gravity correspondence \cite{21}.

Among the parallel, intersecting, and non-intersecting 
configurations of the D-branes the first and the second 
setups have been vastly investigated, while the third one 
has not attracted a considerable attention of the researchers.
In fact, this configuration has some particular and remarkable 
properties. For example, for a system of two 
non-intersecting D-branes the emitted 
closed strings from one of the branes usually 
do not hit to the second brane. Therefore,
a zero or very weak interaction is expected.
For a system of the non-intersecting
perpendicular branes we shall demonstrate 
that this expectation does not occur. 
However, some other properties concerning this configuration 
will be extracted. 

In general, let $d_{\rm or}$ be the 
dimension of the orbifoldy directions, and $d_I$ be
the number of those directions which are perpendicular 
to both branes. For having a setup 
of completely perpendicular branes, i.e. 
branes without any common spatial direction,
there is the following restriction
\begin{eqnarray}
d = 1 + p_1 + p_2 + d_I + d_{\rm or}.
\nonumber
\end{eqnarray}
For the  setups with non-intersecting perpendicular branes 
we should have $d_I \geq 1$, i.e. 
\begin{equation}
p_1 + p_2 \leq d- d_{\rm or}-2.
\end{equation}

Now we consider the following embedding for the 
non-intersecting perpendicular D$p_1$- and D$p_2$-branes
\begin{table}[htp]
\renewcommand*{\arraystretch}{2}
\begin{tabular}{ c | c | c | c | c | c}

$ $ 
& $x^0 $ 
& $ x^1, \ldots , x^{p_1} $
& $ x^{p_1+1}, \ldots , x^{p_1+p_2} $
& $ x^{p_1+p_2+1}, \ldots , x^{d-5} $
& $ x^{d-4}, \ldots , x^{d-1} $ \\
\hline
$D{p_1}:$ & $\times $ & $\times\ldots\times $ & $ $
& $ $ & $ $ \\
\hline
$D{p_2}: $ & $\times$ & $ $ & $\times\ldots\times $ & $ $
& $ $
\end{tabular}
\end{table}

Therefore, for our setup with four orbifoldy directions 
Eq. (3.1) leads to the condition $p_1 + p_2 \leq d-6 $.

The interaction between two D-branes can be conveniently computed by 
the overlap of the two boundary states, corresponding to
the two D-branes, via the closed string propagator. That is, 
$\mathcal{A}=\langle B_1 |D| B_2 \rangle$ 
where the closed string propagator $D$ 
is defined by
$$D=2\alpha'\int_{0}^{\infty}dt~e^{-tH_{\rm closed}}~,$$
and the closed string Hamiltonian is given by 
\begin{equation}
H_{\rm closed}=H_{\rm ghost}+\alpha'p^{\lambda}p_{\lambda}
+2\left(\sum_{n=1}^{\infty}(\alpha_{-n}^{\lambda}
\alpha_{n\lambda}
+\tilde{\alpha}_{-n}^{\lambda}\tilde{\alpha}_{n\lambda})
+\sum_{r=1/2}^{\infty}
(\alpha_{-r}^{a}\alpha^a_{r}
+\tilde{\alpha}_{-r}^{a}\tilde{\alpha}^a_{r})\right)-\frac{d-6}{6}~.
\label{asd}
\end{equation} 
The change of the zero-point 
energy of the Hamiltonian is due to the orbifold projection. 

After lengthy calculations the interaction 
amplitude finds the following form
\begin{eqnarray}
\mathcal{A}&=&\frac{\sqrt{2\pi}T_{p_{1}}T_{p_{2}}\alpha'V_0}
{2(2\pi)^{d-p_1-p_2-5}}
\frac{\prod_{n=1}^{\infty}
\left[\det M^{\dag}_{(n)1} 
\det M_{(n)2}\right]^{-1}}{\sqrt{\det{(U_1/2)}
\det{(U_2/2)}}}
\nonumber\\
&\times& \int_{0}^{\infty}dt\bigg\{e^{(d-8)t/6}
\Big{(}\sqrt{\frac{\pi}{\alpha' t}}\Big{)}^{d_{I_{n}}}
\exp\left( {-\frac{1}
{4\alpha't}\sum_{I_n}{\left(y_{1}^{I_{n}}-y_{2}^{I_{n}}\right)^2}}\right)
\nonumber\\
&\times&
\prod_{I_{c}}\Theta_{3}
\left(\dfrac{y_{1}^{I_{c}}-y_{2}^{I_{c}}}
{2\pi R_{I_{c}}}\vert \dfrac{i\alpha't}{\pi R_{I_{c}}^{2}}
\right)
\frac{\exp\bigg{[}\frac{1}{2}
\bigg{(}\chi^{\dag}
G^{-1}\chi
+\rho^{\dag}R^{-1}
\rho\bigg{)}+\xi\bigg{]}}
{\sqrt{\eta \; \det G\det R}}
\nonumber\\
&\times&
\prod_{n=1}^\infty \bigg{(} 
\det[1-S^{\dag}_{(n)1}S_{(n)2}e^{-4nt}]^{-1}~
(1- e^{-4nt})^{2}(1- e^{-2(2n-1)t})^{-4}\bigg{)}\bigg\},
\label{tg}
\end{eqnarray}
where 
\begin{eqnarray}
\eta \equiv 2t\alpha'+i\alpha'\left[
\mathcal{Q}_{1}^T U_{1}^{-1}+U_{1}^{-1}\mathcal{Q}_{1}
-\mathcal{Q}_{2}^{T}U_{2}^{-1}
-U_{2}^{-1}\mathcal{Q}_{2}\right]_{00},
\nonumber
\end{eqnarray}
and $V_0$ is the common worldvolume of the branes,
i.e. the length of the time direction.
We decomposed the directions 
$\{x^I|I=p_1+p_2+1,\ldots ,d-5\}$, 
which are perpendicular to both branes,
into the compact and non-compact subsets, i.e.
\begin{eqnarray}
\{I\} = \{I_n\} \cup \{I_c\} .
\nonumber
\end{eqnarray}
$d_{I_n}$ is the number of the noncompact directions
$\{x^{I_n}\}$.
The other variables have the following definitions
\begin{eqnarray}
G_{\alpha'_{1}\beta'_{1}}
&=&2t\alpha'\delta_{\alpha'_{1}\beta'_{1}}
+i\alpha'(\mathcal{Q}_{1}^T U_{1}^{-1}
+U_{1}^{-1}\mathcal{Q}_{1})_{\alpha'_{1}\beta'_{1}}
\nonumber\\
&-& 2\alpha'(\mathcal{Q}_{1}^T U_{1}^{-1}
+U_{1}^{-1}\mathcal{Q}_{1})_{\alpha'_{1}0}
(\mathcal{Q}_{1}^T U_{1}^{-1}
+U_{1}^{-1}\mathcal{Q}_{1})_{\beta'_{1}0},
\nonumber\\
R_{\alpha'_{2}\beta'_{2}}
&=&2t\alpha'\delta_{\alpha'_{2}\beta'_{2}}
-i\alpha'(\mathcal{Q}_{2}^T U_{2}^{-1}
+U_{2}^{-1}\mathcal{Q}_{2})_{\alpha'_{2}\beta'_{2}}
\nonumber\\
&-& 2\alpha'(\mathcal{Q}_{2}^T U_{1}^{-1}
+U_{2}^{-1}\mathcal{Q}_{2})_{\alpha'_{2}0}
(\mathcal{Q}_{2}^T U_{2}^{-1}
+U_{2}^{-1}\mathcal{Q}_{2})_{\beta'_{2}0},
\nonumber\\
\mathcal\chi^{\beta'_{1}}&=&-2(U_{1}^{-1}
\mathcal{F}_{1})_{\alpha'_{1}}^{\;\;{\beta'_{1}}}
\ell^{\alpha'_{1}}-y_{2}^{\beta'_{1}}
-4\alpha'(U_{1}^{-1}\mathcal{F}_{1})_{\alpha'_{1}0}
\ell^{\alpha'_{1}}(\mathcal{Q}_{1}^T
U_{1}^{-1}+U_{1}^{-1}\mathcal{Q}_{1})^{\beta'_{1}0}
\nonumber\\
&+&4\alpha'(U_{2}^{-1}\mathcal{F}_{2})_{\alpha'_{2}0}
\ell^{\alpha'_{2}}(\mathcal{Q}_{1}^T
U_{1}^{-1}+U_{1}^{-1}\mathcal{Q}_{1})^{\beta'_{1}0},
\nonumber\\
\mathcal\rho^{\beta'_{2}}&=&-2(U_{2}^{-1}
\mathcal{F}_{2})_{\alpha'_{2}}^{\;\;{\beta'_{2}}}
\ell^{\alpha'_{2}}-y_{1}^{\beta'_{2}}
+4\alpha'(U_{1}^{-1}\mathcal{F}_{1})_{\alpha'_{1}0}
\ell^{\alpha'_{1}}(U_{2}^{-1}\mathcal{Q}_{2}
+\mathcal{Q}_{2}^{T}U_{2}^{-1})^{\beta'_{2}0}
\nonumber\\
&-&4\alpha'(U_{2}^{-1}\mathcal{F}_{2})_{\alpha'_{2}0}
\ell^{\alpha'_{2}}(U_{2}^{-1}\mathcal{Q}_{2}
+\mathcal{Q}_{2}^{T}U_{2}^{-1})^{\beta'_{2}0},
\nonumber\\
\mathbf{\xi}&=& -\frac{1}{\eta}
\bigg{[}2(\mathcal{F}_{1}U_{1}^{-1})_{\alpha'_{1}0}
\ell^{\alpha'_{1}}(\mathcal{F}_{1}
U_{1}^{-1})_{\beta'_{1}0}\ell^{\beta'_{1}}
+2(U_{2}^{-1}\mathcal{F}_{2})_{\alpha'_{2}0}
\ell^{\alpha'_{2}}(U_{2}^{-1}
\mathcal{F}_{2})_{\beta'_{2}0}\ell^{\beta'_{2}}
\nonumber\\
&+&4(\mathcal{F}_{1}U_{1}^{-1})_{\alpha'_{1}0}
\ell^{\alpha'_{1}}(U_{2}^{-1}
\mathcal{F}_{2})_{\beta'_{2}0}\ell^{\beta'_{2}}\bigg{]}~.
\end{eqnarray}
In these variables the primed indices represent the 
spatial directions of the branes, e.g. for the 
D$p_1$-brane there is 
$\{x^{\alpha_1}\}=\{x^0\}\cup \{x^{\alpha'_1}
|\alpha'_1 = 1, 2, \ldots, p_1\}$.
 
We see that the common worldvolume of the branes,
i.e. $V_0$, appeared in the amplitude. For 
the non-intersecting perpendicular branes $V_0$
is the length of the time direction, which is nonzero 
and hence there is an interaction. 
This elaborates that only a portion of 
the interaction is the 
effect of ``emitting closed strings from volume of 
one brane and absorbing them by the volume of the other brane''.
The main part of the interaction is due
to the fact that the worldsheets of the exchanged closed strings 
intersect the worldvolumes of both branes.

The amplitude (3.3) implies that the strength of the interaction 
is exponentially damped by the square distance of the branes. 
In the fourth line the determinant part is contribution of the 
oscillators of the non-orbifoldy directions,
the factor $\prod_{n=1}^\infty(1- e^{-4nt})^{2}$ 
is the ghosts contribution, and 
$\prod_{n=1}^\infty(1- e^{-2(2n-1)t})^{-4}$
comes from the oscillators of the orbifoldy directions.
Note that Eq. (3.3) has the symmetry 
$\mathcal{A}^{\dag}|_{1\leftrightarrow 2}=\mathcal{A}$,
which is induced by the formula $\mathcal{A}=\langle
B_1|D|B_2\rangle$. 

For acquiring the interaction amplitude 
in the non-compact orbifoldized spacetime, in Eqs. (3.3) 
and (3.4) we should replace   
$I_n \rightarrow I$, $d_{I_n} \rightarrow d_I=d-p_1-p_2-5$, 
$\Theta_3 \rightarrow 1$, $\ell^{\alpha'_1} \rightarrow 0$
and $\ell^{\alpha'_2} \rightarrow 0$.
\subsection{Interaction of the distant branes}

Behavior of the interaction amplitude 
for the distant branes represents the long-range
force of the theory. That is, after an 
enough long time the massless closed string states possess   
a dominant contribution on the interaction, while the 
contributions of all massive states, except 
the tachyon state, are drastically damped.
Therefore, in the dimension $d = 26$, we introduce the limit
$t \rightarrow \infty $ on the oscillating 
part of the general amplitude, i.e. on the last line of Eq. (3.3).
Since in this dimension the states of the metric, antisymmetric 
tensor and dilaton are
massless they have zero winding and zero 
momentum numbers. Thus, we receive 
\begin{eqnarray}
\mathcal{A}_{\rm long-range}
&=&\frac{\sqrt{2\pi}T_{p_{1}}T_{p_{2}}\alpha'V_0}
{2(2\pi)^{d-p_1-p_2-5}}
\frac{\prod_{n=1}^{\infty}
\left[\det M^{\dag}_{(n)1} 
\det M_{(n)2}\right]^{-1}}{\sqrt{\det{(U_1/2)}
\det{(U_2/2)}}}
\nonumber\\
\nonumber\\
&\times &
\int_{0}^{\infty}dt\bigg\{\Big{(}\sqrt{\frac{\pi}{\alpha' t}}
\Big{)}^{d_{I_{n}}}
\exp\left( {-\frac{1}
{4\alpha't}\sum_{I_n}{\left(y_{1}^{I_{n}}-y_{2}^{I_{n}}\right)^2}}\right)
\nonumber\\
&\times&
\prod_{I_{c}}\Theta_{3}
\left(\dfrac{y_1^{I_{c}}-y_2^{I_{c}}}
{2\pi R_{I_{c}}}\big{|} \dfrac{i\alpha't}
{\pi R_{I_{c}}^{2}}\right)
\frac{\exp\bigg{[}\frac{1}{2}\left( y^T_2
G^{-1}y_2+y^T_1 R^{-1}y_1 \right)
\bigg{]}}{\sqrt{ \eta \;\det G \det R}}
\nonumber\\
\nonumber\\
&\times&
\lim_{t \to \infty}\left(e^{3t}+
{\rm Tr}(S^{\dag}_{(n=1)1}S_{(n=1)2})
e^{-t}\right)\bigg\}~.
\label{tg}
\end{eqnarray}

We observe that the 
contribution of the massless states, i.e. the 
gravitation, dilaton and Kalb-Ramond,
vanishes. Therefore, the orbifold projection 
extremely eliminates the long-range force. 
Precisely, this is an effect of the constant term 
of the closed string Hamiltonian (3.2) which is imposed 
by the orbifold projection. 
Note that the untwisted sector of the theory possesses 
the long-range force. Hence, the total amplitude 
which comes from the both twisted and untwisted 
sectors contains a non-vanishing long-range force.
However, according to the negative mass squared of the 
tachyon, the divergent part of Eq. (3.5) elaborates 
exchange of the tachyonic state. Due 
to the orbifold projection this divergence 
is unlike the conventional case.

\section{Conclusions and summary}

We obtained the boundary state corresponding to a 
dynamical fractional-wrapped D$p$-brane
in the following backgrounds: the Kalb-Ramond field, a $U(1)$ 
gauge potential and a tachyon field. 
The brane lives in an orbifold-toroidal spacetime
$T^{n} \times \mathbb{R}^{1, d-n-5} 
\times\mathbb{C}^{2}/\mathbb{Z}_{2}$. 
We observe that presence of the tachyon field  
prominently affects the winding of the closed strings around
the wrapped directions of the brane.

The emitted closed string 
acquires a momentum along the brane worldvolume,
which is unlike the conventional case.
This momentum depends on the position of the 
closed string and its winding numbers around the compact 
directions of the brane. This peculiar result 
demonstrates that the toroidal compactification, 
background fields, linear 
and angular motions of the brane, impose a 
potential on the emitted closed strings.

The amplitude of the interaction for 
two non-intersecting perpendicular  
fractional-wrapped D$p_1$- and D$p_2$-branes, in the 
foregoning setup, was calculated. Since the amplitude 
is nonzero we conclude 
that the worldvolumes of the branes and the worldsheets
of the exchanged closed string play 
the main role for the interaction.
Precisely, for a system of non-intersecting branes 
the worldsheet of any exchanged  
closed string intersects the worldvolumes of both branes,
and hence interaction takes place. However,
the various parameters, i.e., the elements of the 
Kalb-Ramond and 
tachyon matrices, the angular and linear velocities of the branes,
the dimensions of the spacetime and branes, the
orbifoldy directions, the closed string 
winding and momentum numbers, the coordinates 
of the branes location, and the radii of 
compactification, give a general feature to the interaction 
amplitude. The strength of the interaction can be accurately
adjusted by these parameters.

From the total amplitude the interaction concerning 
the exchange of the massless states and tachyon state 
was extracted. In the $26$-dimensional spacetime 
the contribution of the massless states 
vanishes. This is an unconventional and 
marvelous effect which was imposed by 
the orbifold projection.
In the dimension $d =32$ we receive a long-range force.
In this dimension the divergence part of 
this interaction, due to the tachyon 
exchange, finds the conventional  
form $e^{4t} \rightarrow \infty$. Generally,
for a given dimension of the orbifold 
one can manipulate the spacetime dimension 
to obtain a large distance amplitude of the 
branes: in the damping form, 
in the usual form and or in the divergent form.

  
\end{document}